\documentclass[aps,prd,groupedaddress,showpacs]{revtex4}
\usepackage{epsf,epsfig,graphics,graphicx,amssymb,amsthm}
\begin{document}

\title{Effects of a pre-inflation radiation-dominated epoch to CMB anisotropy}

\author{I-Chin Wang$^1$ and Kin-Wang Ng$^2$}
\affiliation{
$^1$Department of Physics, National Cheng Kung University, Tainan, Taiwan 701, R.O.C.\\
$^2$Institute of Physics, Academia Sinica, Taipei, Taiwan 115, R.O.C.\\
}

\date{October 12, 2007}

\begin{abstract}
We consider that the pre-inflation era is radiation-dominated,
transiting smoothly to the inflationary era. We work out in detail
the dynamics of inflaton fluctuations across the phase transition
and the proper choices of initial vacuum states. It is found that
this phase transition can suppress long-wavelength quantum
fluctuations of inflaton. This may attribute to the large-scale
CMB anisotropy a lower power than predicted in the standard
$\Lambda$CDM model. In constraining this transitional effect by
WMAP anisotropy data, we use the WMAP best-fit scale-invariant
$\Lambda$CDM model with the density power spectrum replaced by the
one found in this work. We find that the transition occurs at
least about $10$ e-folds before the comoving scales comparable to
our present horizon size cross the Hubble radius during inflation.
\end{abstract}

\pacs{98.80.Cq, 98.70.Vc, 98.80.Es}
\maketitle

\section{Introduction}

The inflationary model can explain the observed spatially flat and
homogenous Universe. Moreover, quantum fluctuations during
inflation can give rise to primordial density fluctuations with an
almost scale-invariant power spectrum, which is consistent with
the recent WMAP data on cosmic microwave background
anisotropies~\cite{spergel06}. Among various cosmological models,
the $\Lambda$CDM model is a quite good one -- it can fit CMB data
well. However, the CMB anisotropy measurements made by WMAP have
shown that the amplitude of the quadrupole is lower than expected
for the $\Lambda$CDM model, although the statistical significance
of such an anomaly is not large~\cite{spergel06}. Recently, there
have been many proposed solutions for this anomaly, mostly based
on some new ingredients in the generating process of density
perturbation~\cite{fang}.

A general assumption of the $\Lambda$CDM model is that it
considers only a vacuum-dominated inflationary epoch, and that it
does not consider the possibility of a pre-inflation era. For
example, a radiation-dominated era has taken place before
inflation. In fact, this phase transition has been proposed in the
context of spontaneous symmetry breaking to provide with a
mechanism for inflation~\cite{kolb90}. In this paper, we will take
into account the effect of a pre-inflation radiation-dominated
era. As expected, if inflation lasts very long such that the
present Universe only occupies a tiny portion of the inflated
region, the effect of the phase transition will be negligible.
However, if it happens that the present Universe is just
comparable to the size of the inflated region, the pre-radiation
era may leave imprints on large-scale structures of the Universe.
Here we will study the effect of the phase transition to the
large-scale CMB anisotropy. Recently, this effect has been
discussed by Powell and Kinney~\cite{PK06}, who considered
large-wavelength inflaton fluctuations in the pre-inflation
radiation-dominated era and found a suppression of power on large
scales as a result of the choices of vacuum states. (See also
Ref.~\cite{kaloper} for a similar consideration.) In addition,
Contaldi {\it et al.} in Ref.~\cite{fang} have found a similar
effect caused by a pre-inflation epoch that is dominated by the
kinetic energy of the inflation. However, the issues about the
dynamics of inflaton fluctuations across the phase transition and
the proper choices of initial vacuum states have not been fully
investigated. Here, we will study these issues in detail.

Let us simply consider that the Universe has two components during
the phase transition -- one is radiation and the other is vacuum
energy. We assume that the pre-inflationary Universe is in a
radiation-dominated phase and then the Universe evolves to a
vacuum-energy-dominated or de Sitter phase. Although the solution
for this transition has been obtained for many
years~\cite{VF82,S82,L82,ENO88}, there is no attempt in varying the
amounts of radiation and vacuum energies of the Universe to see
the effect on CMB anisotropy. This serves a way to test the
existence of an unknown pre-inflation phase. Also, it provides a
constraint on the period of inflation which needs to be consistent
with the observed CMB anisotropy spectrum under such a
pre-inflation condition.

In this paper, we use Vilenkin and Ford's transition
solution~\cite{VF82} to compute the power spectrum of quantum
fluctuations of the inflaton field. Since the pre-inflation phase is
radiation-dominated, we use the radiation-dominated solution to
determine the boundary conditions. This allows us to set the form of the
seed for the quantum fluctuations before the de Sitter phase begins.
By introducing this pre-inflation initial condition, we can
calculate the power spectrum of the horizon-crossing modes near the
transition region. Then we input the power spectrum to the CMBFAST
numerical code~\cite{seljak96} to generate the CMB anisotropy power
spectrum. We vary the amounts of radiation and vacuum energies and
tune the duration of inflation to obtain different results. In order
to compare the results with the $\Lambda$CDM model, we do the
chi-square fitting of the generated CMB anisotropy power spectra
to the WMAP data and calculate the chi-square values.

In next section, we will derive the mode equation of the quantum
fluctuation during the phase transition. The numerical results for
the mode equation and the respective CMB anisotropy power spectra
will be given in Sec.~\ref{numerical}. Sec.~\ref{conclusion} is our
conclusion.

\section{The Model}
\label{model}

We consider a flat Robertson-Walker metric, with a scale factor
$a(t)$ :
\begin{equation}\label{metric}
ds^2=g_{\mu\nu} dy^{\mu} dy^{\nu}= dt^2 -a^2(t) d{\vec y}^2 \;.
\end{equation}
The total energy-momentum tensor is given by
\begin{equation}\label{emtensor}
T_{\mu\nu}=T^{(rad)}_{\mu\nu}+V_0 g_{\mu\nu} \;,
\end{equation}
where $T^{(rad)}_{\mu\nu}$ is the trace-free energy momentum
tensor for radiation and $V_0$ is constant vacuum energy. Hence,
the Einstein evolution equation is
\begin{equation}
H^2\equiv
\left(\frac{1}{a}\frac{da}{dt}\right)^2=\frac{\rho}{3M_{Pl}^2}\;\;,
\label{aeq}
\end{equation}
where $M_{Pl}=(8\pi G)^{-1/2}=2.436\times 10^{18}\;{\rm GeV}$ is the
reduced Planck mass and $\rho$ is given by
\begin{equation}
\rho=3M_{G}^4\left(\frac{A}{a^4}+B\right)\;, \label{rhoeq}
\end{equation}
with $V_0=3B M_{G}^4$ of order of the grand unification energy
scale. $A$ and $B$ are dimensionless constant parameters. Defining
$\Lambda^2\equiv M_{G}^4/M_{Pl}^2$, the solution of
Eqs.~(\ref{aeq}) and (\ref{rhoeq}) is found as~\cite{VF82}
\begin{equation}
a(t)=\left(\frac{A}{B}\right)^{\frac{1}{4}}\left[\sinh\left(2\sqrt{B}\Lambda
t\right)\right]^{\frac{1}{2}}\;. \label{a(t)}
\end{equation}
The asymptotic form in the radiation-dominated phase is
\begin{equation}
 \Lambda t\ll\frac{1}{2} B^{-\frac{1}{2}}  \; \;, \;\;
 a(t)\sim\sqrt{2}A^{\frac{1}{4}}(\Lambda t)^{\frac{1}{2}} \;.
\end{equation}
The asymptotic form in the inflationary phase is
\begin{equation}
\Lambda t\gg\frac{1}{2} B^{-\frac{1}{2}} \; \;, \;\;
 a(t)\sim 2^{-\frac{1}{2}}\left(\frac{A}{B}\right)^{\frac{1}{4}}e^{\sqrt{B}\Lambda t} \;.
\end{equation}

To study the transition from the pre-inflation
radiation-dominated phase to inflation, we put Eq.(\ref{a(t)})
into the equation of motion for the inflaton quantum fluctuation.
To a good approximation, we consider the zero-mass case. The
equation of motion for the quantum fluctuation of inflaton is
given by
\begin{equation}
\Box\varphi(t,{\vec y})=0 , \label{phieq}
\end{equation}
where
\begin{equation}
\Box=-\frac{1}{\sqrt{-g}}\partial_{\mu}(\sqrt{-g}g^{\mu\nu}
\partial_{\nu}).
\end{equation}
The Fourier component of $\varphi(t,{\vec y})$ is given by
\begin{equation}
\phi_k(t)=\int\varphi(t,{\vec y})e^{i{\vec k}\cdot{\vec y}}d{\vec
y}. \label{phik}
\end{equation}
Put Eq.~(\ref{phik}) into Eq.~(\ref{phieq}), then Eq.~(\ref{phieq}) becomes
\begin{equation}
\ddot{\phi}_k(t)+3\frac{\dot{a}}{a}\dot{\phi}_k(t)+\left(\frac{k^2}{a^2}\right){\phi}_k(t)=0
\;\;,\;\;\dot{\phi}_k(t)\equiv \frac{d\phi_k(t)}{dt} \;.
\label{phikeq}
\end{equation}

To solve Eq.~(\ref{phikeq}) with $a(t)$ given by Eq.~(\ref{a(t)}),
we change the derivative of an independent variable from the time
$t$ to the scale factor $a(t)$. As we want to study the effect of
the transition around the critical point $a_{c}$, where $a_{c}$ is
given by $A/a_{c}^4=B$, we rewrite Eq.~(\ref{phikeq}) in terms of
$x\equiv a-a_{c}$. Then, Eq.~(\ref{phikeq}) becomes
\begin{widetext}
\begin{equation}
\left[B(x+a_{c})^{4}+A\right]\phi''_k(x)+\left[4B(x+a_{c})^{3}+
\frac{2A}{x+a_{c}}\right]\phi'_k(x)+k^2\phi_k(x)=0
\;\;,\;\;\;\phi'_k(x)\equiv \frac{d\phi_k(x)}{dx}\;\;\;.
\label{phixeq}
\end{equation}
\end{widetext}
Note that when $a(t)$ is small, the Universe is in
radiation-dominated era. Therefore, we choose the boundary
conditions for Eq.~(\ref{phixeq}) as
\begin{eqnarray}
\phi_k(x)=\frac{1}{a}\frac{1}{\sqrt{2k}} e^{ika/\sqrt{A}}\;\;\;,
\label{bc1} \\
\phi'_k(x)=\left[-\frac{1}{\sqrt{2k}
a^2}+\frac{i\sqrt{k}}{\sqrt{2A} a}\right] e^{ika/\sqrt{A}}\;\;\;.
\label{bc2}
\end{eqnarray}
Solving Eq.~(\ref{phixeq}) with above boundary conditions gives
the solution for a particular $k$ mode. The relation $k=aH$
determines the value of $k$ for the mode that crosses the horizon
at $a=a_k$, where $H$ is the Hubble parameter given by
Eq.~(\ref{aeq}). This gives from Eq.~(\ref{aeq}) that
\begin{equation}
{k\over \Lambda} = \left(\frac{A}{a_k^2}+B a_k^2\right)^{1\over2}.
\end{equation}
Since we are concerned with the $k$-modes that leave the horizon
shortly after inflation begins, we will consider only the
solutions with $k\geq k_{c}$, where $k_{c}=\Lambda
\sqrt{2B}a_{c}$. The power spectrum of the quantum fluctuations of
the inflation is then given by
\begin{equation}
P_{k}^{1/2}\propto k^{3/2}\phi_k(x_k)\;, \label{ps}
\end{equation}
where $x_k=a_k-a_c$.
In the following section, we compute power spectra $P_{k}^{1/2}$
produced when and after inflation begins ($x\geq 0$) by
numerically solving Eq.~(\ref{phixeq}) for different values of $A$
and $B$. Then, we put these results into the CMBFAST
code~\cite{seljak96} to generate the CMB anisotropy power spectra.

\section{Numerical Results}
\label{numerical}

So far, we have not specified the values of
the parameters $A$ and $B$. Before doing the
numerical calculation, let us assess the possible range of the
values of $A$ and $B$.

From Eq.~(\ref{rhoeq}), the vacuum energy that drives inflation is
given by
\begin{equation}
V_0=3 M_{G}^4 B\;.
\end{equation}
Here we consider $V_0$ as the potential energy of an inflaton
potential $V(\varphi)$ during inflation. The measurements of the
CMB anisotropy made by WMAP has put a constraint on $V(\varphi)$
given by~\cite{peiris06}
\begin{equation}
V^{1\over4}= 0.0265\epsilon^{1\over4}M_{Pl},
\end{equation}
where the slow-roll parameter
\begin{equation}
\epsilon\equiv (M_{Pl}^2/2)(V'/V)^2 < 0.033
\end{equation}
at 95\% confidence level. Therefore, we have $V_0\simeq V$ and thus
\begin{equation}
B^{\frac{1}{4}}< 0.0086 M_{Pl}/M_G.
\end{equation}
If we take $M_G=2.1\times 10^{16}\;{\rm GeV}$, then $B<1$.

In the following, we choose $B=1$ and $B=0.1$ to calculate the
power spectra of inflaton fluctuations $P_k$ in Eq.~(\ref{ps}).
Although we have introduced the slow-roll parameter $\epsilon$, we
will assume a pure de Sitter space during inflation and neglect
the small slow-roll corrections to the density power spectrum.
This would not affect our interest here since the effect of a
pre-inflation radiation-dominated epoch is mainly on large angular
scales. Then, we use the resulting density power spectrum and the
other cosmological parameters in the WMAP best-fit scale-invariant
$\Lambda$CDM model (i.e. the density power spectrum has a scalar
spectral index $n_s=1$)~\cite{spergel06} to generate the CMB
anisotropy power spectra. Note that while the observation gives
the constraint on $B$, the value of $A$ is undetermined. In fact,
different values of $A$ introduce different boundary conditions as
shown in Eqs.~(\ref{bc1}) and (\ref{bc2}). Therefore, we pick a
certain value of $A$ to see how the value affects the CMB
anisotropy power spectrum and study under what conditions the
results are compatible with present CMB observations. In order to
give a quantitative comparison with the CMB data, we use the
chi-square fitting,
$\chi^2=\sum_{\imath,\jmath}(D^{b}_{\imath}-T^{b}_{\imath})
C^{-1}_{\imath,\jmath} (D^{b}_{\jmath}-T^{b}_{\jmath})$, and the
chi-square value as a measure, where $D^{b}_{\imath}$ is the
measured $\imath$th band-power, $T^{b}_{\jmath}$ is the
theoretical value, and $C_{\imath,\jmath}$ is the width of the
error bar in measurement.

\begin{figure}[htp]
\centering
\includegraphics[width=0.62\textwidth]{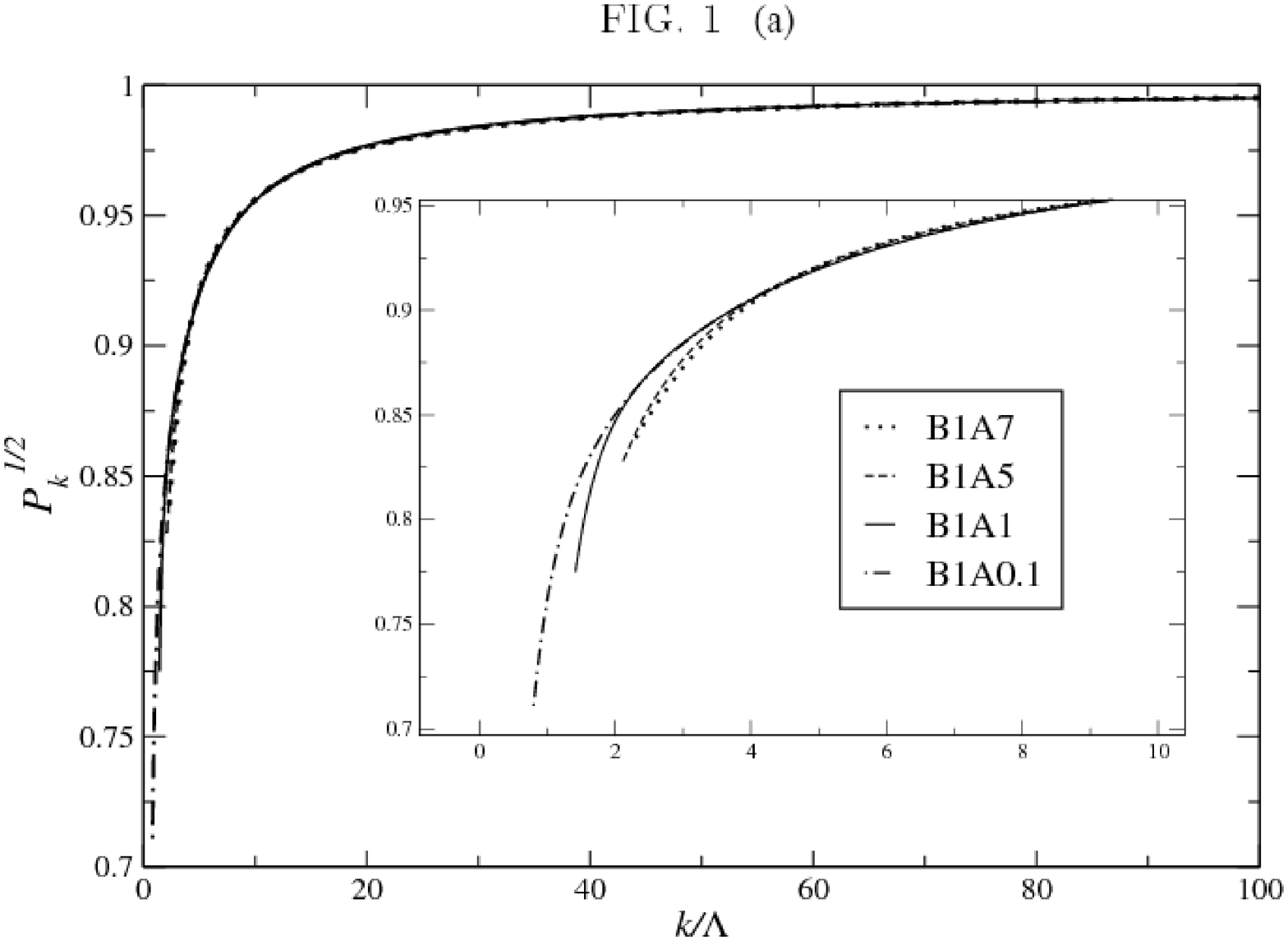}
\includegraphics[width=0.62\textwidth]{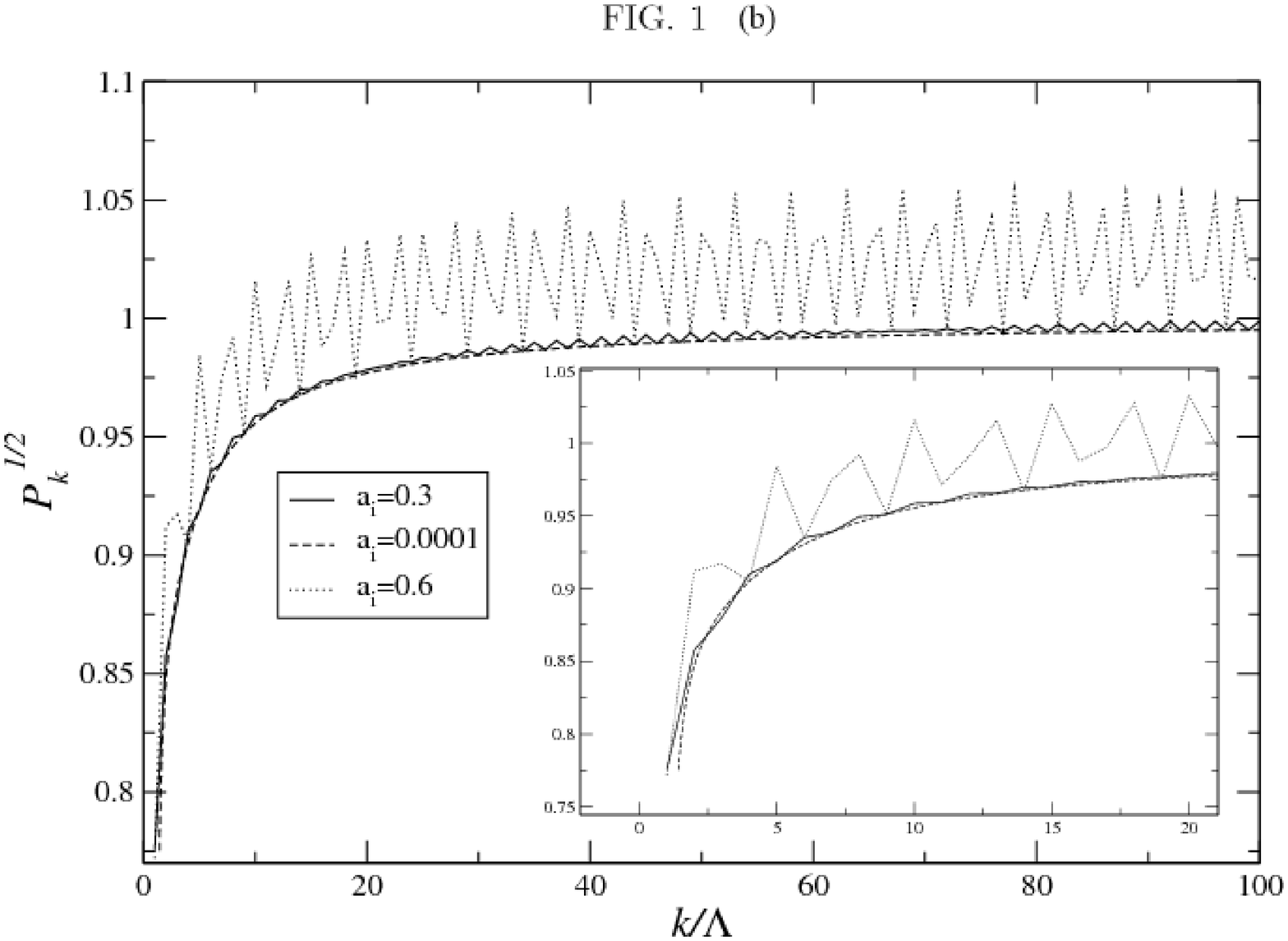}
\caption{ (a) Normalized power spectra of inflaton fluctuations
with $A= 7, 5, 1, 0.1$ for a fixed $B=1$. Inserted panel shows the
spectra for small $k$. All curves start with $k=k_c$ corresponding
to the $k$-mode that leaves the horizon at the start of inflation.
(b) Normalized power spectra of inflaton fluctuations with initial
$a_{i}=0.001, 0.3, 0.6$ for the case with $B=A=1$. The phase
transition point in this case is $a_{c}=1$.} \label{fig1}
\end{figure}

We present the following two cases:

(1) $B=1$ with $A= 7, 5, 1, 0.1$. In this case, the transition
takes place at $a_c=1.63-0.56$ with $k_c/\Lambda=2.30-0.80$.
Figure~\ref{fig1}(a) shows the power spectrum of inflaton
fluctuations $P_k^{1/2}$ for each set of $B$ and $A$. It is
apparent that the large-scale or small-$k$ power is suppressed by
the presence of the pre-inflation radiation-dominated epoch. This
suppression is similar to that found in Ref.~\cite{PK06}. However,
there are two main differences. For large-$k$ modes, we have
assumed a standard inflationary scale-invariant power spectrum
with $n_s=1$ while they have taken the spectrum with $n_s=0.951$.
For small-k modes, our spectrum increases monotonically with $k$
whereas their spectrum undergoes oscillations; however, both
spectra rapidly approach the inflationary spectrum. This
oscillatory behavior, also found by Contaldi {\it et al.} in
Ref.~\cite{fang} (the result from an analytic computation as shown
in Fig.~1 of their paper) and Enqvist {\it et al.} in
Ref.~\cite{ENO88}, is actually an artifact resulted from the
discontinuity of Ricci curvature at the abrupt phase transition.
This artificial particle production can be avoided by considering
for example a $C^\infty$ function for the scale factor
$a(t)$~\cite{chung}, as we have worked with in Eq.(\ref{a(t)}).

Besides, we must be very careful about choosing the initial point
$a_{i}$ while doing the numerical calculation. For example, in our
case with $B=A=1$, the phase transition point is at $a_c=1$. We
have obtained a smooth power spectrum (see Fig.~\ref{fig1}(a)) by
choosing the initial $a_{i}$ to be $a_{i}=0.0001$ (which means
that it is in the very early Universe; therefore, it is purely
radiation-dominated and also very far from the phase transition
point $a_c=1$). But, if we tune the initial $a_{i}$ to be close to
the phase transition point, say $a_{i}=0.3$ , the oscillation
begins to show up. And if we tune the initial $a_{i}$ closer to
the phase transition point, for example, $a_{i}=0.6$ , the
oscillation becomes larger than that in the $a_{i}=0.3$ case, as
shown in Fig.~\ref{fig1}(b). This test explains that if the
initial $a_{i}$ is not in the proper region (proper region means
the radiation-dominated region and that is the right region for
the radiation-dominated solution in Eq.~(\ref{bc1}), while
improper region means the phase transition region which is not a
radiation-dominated region and therefore the solution for
$\phi_{k}(x)$ is not Eq.~(\ref{bc1}) anymore). This improper
choice for the initial $a_{i}$ indeed leads to an oscillating
power spectrum. However, if the initial $a_{i}$ is carefully
chosen in the proper region, it does not lead to an oscillating
power spectrum. Note that here we have used the scale factor $a$
as the evolution variable. If we have used the conformal time, we
would have obtained an oscillating power spectrum which approaches
the flat inflationary spectrum at large $k$. This explains why the
power spectrum obtained by Contaldi {\it et al.} in
Ref.~\cite{fang} (see Fig.~2 of their paper) is oscillating
although it is obtained from an exact numerical evolution. In
their paper, they have indeed used an improper choice of the
initial condition which is too close to the phase transition.
(They have chosen an initial kinetic energy of the inflaton only
about 100 times of the potential energy. Since the kinetic energy
of the inflaton decreases as $a^{-6}$, both energies rapidly
become comparable. To obtain a correct power spectrum, they should
have used a much larger initial kinetic energy.)

The power suppression would affect the CMB anisotropy power
spectrum. For a spectrum $P_k^{1/2}$, we choose a value of
$z\equiv k_{0.05}/\Lambda$ corresponding to the physical scale of
$0.05\textrm{Mpc}^{-1}$ and then input this $P_k^{1/2}$ into the
CMBFAST code to generate the CMB anisotropy power spectrum $C_l$.
This value of $z$ also corresponds to an exponential expansion
with the number of e-folds from the start of inflation given by
\begin{equation}
N_z=\ln (a_k/a_c) \simeq \ln z.
\end{equation}
The results are shown in Fig.~\ref{fig2}. Since the measured
low-$l$ multipoles of the CMB anisotropy power spectrum are
slightly different in the three-year WMAP data
(WMAP3)~\cite{wmap3} from the one-year data (WMAP1)~\cite{wmap1},
we compare the chi-square fitting by using WMAP3 data with that
with WMAP1 data. The results are summarized in
Tables~\ref{B1wmap3} and \ref{B1wmap1}.

The $\chi^2$ value of the $\Lambda$CDM model with a
scale-invariant power spectrum using WMAP3 data is $47.09$. From
Table~\ref{B1wmap3}, we can see that the $\chi^2$ value for the
model decreases to about $48$ as $z$ increases to $3\times 10^4$,
corresponding to $N_z\simeq 10$ e-folds after the start of
inflation. This can be reflected, for example in Fig.~\ref{fig2}b,
by those five curves for $z=120, 300, 500, 1000, 3000$ rising in
turn from down to top. As we further increase the value of $z$,
the resulting $C_l$ approaches to that of the $\Lambda$CDM model
with a scale-invariant power spectrum. Thus, it can be concluded
that the WMAP3 data does not prefer the presence of a
pre-inflation radiation-dominated epoch, although it is not
excluded. If we assume that the comoving scales comparable to our
present horizon size cross the Hubble radius during inflation when
there are $N_{cmb}\simeq 60$ e-folds from the end of
inflation~\cite{liddle03}, then the radiation-dominated epoch
would take place at least $N=N_z+N_{cmb}\simeq 70$ e-folds from
the end of inflation.

The case is different when using WMAP1 data since the measured
one-year low-$l$ multipoles are slightly smaller than the
three-year results. The $\chi^2$ value of the $\Lambda$CDM model
with a scale-invariant power spectrum using WMAP1 data is $64.99$.
From Table~\ref{B1wmap1}, we can see that the $\chi^2$ values for
all other models are slightly lower than this value at $z\sim
10^4$, corresponding to $N_z\simeq 9$ e-folds after the start of
inflation. Therefore, the WMAP1 data would, albeit a rather low
significance, indicate that the universe would be actually
predominated by radiation when $N=N_z+N_{cmb}\simeq 69$ e-folds
from the end of inflation.

(2) $B=0.1$ with $A= 7, 5, 1, 0.1$. The transition takes place at
$a_c=2.89-1.00$ with $k_c/\Lambda=1.29-0.45$. We do not show the
power spectrum of inflaton fluctuations $P_k$ for this case which
is indeed quite similar to the previous case. Figure~\ref{fig3}
shows the $C_l$ power spectra and their $\chi^2$ values are given
in Tables~\ref{B0.1wmap3} and \ref{B0.1wmap1}. In
Table~\ref{B0.1wmap3}, the $\chi^2$ value decreases rapidly to
about $47$ at $z\sim 3000$; however, at large $z$ all models
approach to the $\Lambda$CDM model with a scale-invariant power
spectrum. In Table~\ref{B0.1wmap1}, the $\chi^2$ values for all
the models reach a minimum at $z\sim 1000$, which is lower than
the $\chi^2$ value for the $\Lambda$CDM model with a
scale-invariant power spectrum. This corresponds to an
inflationary expansion with e-folds of $N_z\simeq 7$ after the
start of inflation. Because the $\chi^2$ values decrease with $z$
in the $B=0.1$ case faster than the $B=1$ case, the anisotropy
power spectra are closer to each other and the $B=0.1$ case
reaches the standard $\Lambda$CDM model at a smaller $z$ value. It
means that inflation in the $B=1$ case lasts longer than the
$B=0.1$ case in order to be consistent with the CMB anisotropy
power spectrum observed today. We have assumed that inflation is
driven by a nearly flat potential and thus neglected any spectral
index. As we can see in Fig.~\ref{fig3} there is a certain
degeneracy in the CMB power spectrum between the variation of the
spectral index and the existence of a pre-inflation
radiation-dominated era, so a joint likelihood fitting of relevant
cosmological parameters to WMAP data should be carried out before
drawing any conclusion about pre-inflation physics.

\section{Conclusion}
\label{conclusion}

We have shown that the existence of a pre-inflation
radiation-dominated phase can affect the CMB anisotropy power
spectrum. This transition can give a low quadrupole value by
adjusting the values for $A$, $B$, and $z$. The smaller is the
value of $z$, the closer is the comoving scales of the present
Universe to the transition when they cross out the horizon during
inflation. This manifests as rising up or down for those low-$l$
$C_l$ of the $z=120$ spectra in Figs.~\ref{fig2} and \ref{fig3}.
Meanwhile, the high-$l$ $C_l$, for example, in Fig.~\ref{fig2}d is
significantly lower than the observed values. So, it is unlikely
that the inflationary era lasts for only $60$ e-folds. As
expected, the effect of the radiation phase becomes weaker as the
value of $z$ increases. From the three-year WMAP data, using the
WMAP best-fit $\Lambda$CDM model parameters, we have found that
the phase transition occurs at least $70$ e-folds from the end of
inflation. Interestingly, the one-year WMAP data indeed prefers
very slightly the pre-inflation radiation-dominated phase taking
place as early as about $70$ e-folds from the end of inflation.
Future CMB observations will provide more severe constraints on
this transitional effect, and a joint likelihood fitting to CMB
data for studying the degeneracy between the addition of a
pre-inflation radiation-dominated era and other relevant
cosmological parameters will certainly be necessary.

\begin{acknowledgments}
This work was supported in part by the National Science Council,
Taiwan, ROC under the Grant NSC 95-2112-M-001-052-MY3.
\end{acknowledgments}

\begin{table}[htp]
\centering
\begin{tabular}{|@{\hspace{0.3cm}}c@{\hspace{0.3cm}}|@{\hspace{0.3cm}}c@{\hspace{0.3cm}}|@{\hspace{0.3cm}}c@{\hspace{0.3cm}}|@{\hspace{0.3cm}}c@{\hspace{0.3cm}}|
@{\hspace{0.3cm}}c@{\hspace{0.3cm}}|@{\hspace{0.3cm}}c@{\hspace{0.3cm}}
|@{\hspace{0.3cm}}c@{\hspace{0.3cm}}|@{\hspace{0.3cm}}c@{\hspace{0.3cm}}
|} \hline
$ $ & $z=120$ & $z=300$ & $z=500$ & $z=10^3$ & $z=3\times10^3$ & $z=10^4$ & $z=3\times10^4$  \\
\hline
$\chi^2(B1A7)$ & $997$ & $367.2$ & $211.3$ & $103$ & $57.91$ & $49.27$ & $47.70$   \\
\hline
$\chi^2(B1A5)$ & $1299$ & $443$ & $217$ & $102$ & $57.43$ & $49.16$ & $47.67$   \\
\hline
$\chi^2(B1A1)$ & $3548$ & $449.4$ & $209.6$ & $98.37$ & $56.12$ & $48.87$ & $47.59$   \\
\hline
$\chi^2(B1A0.1)$ & $1574$ & $411.6$ & $208.8$ & $99$ & $56.54$ & $48.96$ & $47.61$   \\
\hline
\end{tabular}
\caption{Using WMAP3 data to the chi-square fitting of the CMB
anisotropy power spectra for $B=1$ and $A=7, 5, 1, 0.1$ with
different $z$ values. Note that the chi-square value of the
$\Lambda$CDM model with a scale-invariant power spectrum is
$47.09$.} \label{B1wmap3}
\end{table}

\begin{table}[htp]
\centering
\begin{tabular}{|@{\hspace{0.3cm}}c@{\hspace{0.3cm}}|@{\hspace{0.3cm}}c@{\hspace{0.3cm}}|@{\hspace{0.3cm}}c@{\hspace{0.3cm}}|@{\hspace{0.3cm}}c@{\hspace{0.3cm}}|
@{\hspace{0.3cm}}c@{\hspace{0.3cm}}|@{\hspace{0.3cm}}c@{\hspace{0.3cm}}
|@{\hspace{0.3cm}}c@{\hspace{0.3cm}}|@{\hspace{0.3cm}}c@{\hspace{0.3cm}}
|} \hline
$ $ & $z=120$ & $z=300$ & $z=500$ & $z=10^3$ & $z=3\times10^3$ & $z=10^4$ & $z=3\times10^4$  \\
\hline
$\chi^2(B1A7)$ & $639.50$ & $236.37$ & $146$ & $85.45$ & $65.93$ & $64.47$ & $64.72$   \\
\hline
$\chi^2(B1A5)$ & $822.23$ & $280.6$ & $147.16$ & $84.83$ & $65.81$ & $64.48$ & $64.74$   \\
\hline
$\chi^2(B1A1)$ & $2238$ & $283.35$ & $ 142.47$ & $82.90$ & $65.44$ & $64.50$ & $64.77$   \\
\hline
$\chi^2(B1A0.1)$ & $992$ & $261.4$ & $142.2$ & $83.51$ & $65.55$ & $64.49$ & $64.75$   \\
\hline
\end{tabular}
\caption{The same as in Table~\ref{B1wmap3} but using WMAP1 data.
Note that the chi-square value of the $\Lambda$CDM model with a
scale-invariant power spectrum is $64.99$.} \label{B1wmap1}
\end{table}

\begin{table}[htp]
\centering
\begin{tabular}{|@{\hspace{0.3cm}}c@{\hspace{0.3cm}}|@{\hspace{0.3cm}}c@{\hspace{0.3cm}}|@{\hspace{0.3cm}}c@{\hspace{0.3cm}}|@{\hspace{0.3cm}}c@{\hspace{0.3cm}}|
@{\hspace{0.3cm}}c@{\hspace{0.3cm}}|@{\hspace{0.3cm}}c@{\hspace{0.3cm}}
|@{\hspace{0.3cm}}c@{\hspace{0.3cm}}|@{\hspace{0.3cm}}c@{\hspace{0.3cm}}
|} \hline
$ $ & $z=120$ & $z=300$ & $z=500$ & $z=10^3$ & $z=3\times10^3$ & $z=10^4$ & $z=10^5$  \\
\hline
$\chi^2(B0.1A7)$ & $59$ & $56.88$ & $52.82$ & $49.83$ & $47.97$ & $47.43$ & $47.11$   \\
\hline
$\chi^2(B0.1A5)$ & $67.3$ & $56.9$ & $52.59$ & $49.66$ & $47.89$ & $47.32$ & $47.11$   \\
\hline
$\chi^2(B0.1A1)$ & $80.08$ & $56.19$ & $51.78$ & $49.08$ & $47.67$ & $47.23$ & $47.09$   \\
\hline
$\chi^2(B0.1A0.1)$ & $77.98$ & $55.18$ & $51$ & $48.63$ & $47.49$ & $47.2$ & $47.08$   \\
\hline
\end{tabular}
\caption{$\chi^2$ values for $B=0.1$ and $A=7, 5, 1, 0.1$, using
WMAP3 data.} \label{B0.1wmap3}
\end{table}

\begin{table}[htp]
\centering
\begin{tabular}{|@{\hspace{0.3cm}}c@{\hspace{0.3cm}}|@{\hspace{0.3cm}}c@{\hspace{0.3cm}}|@{\hspace{0.3cm}}c@{\hspace{0.3cm}}|@{\hspace{0.3cm}}c@{\hspace{0.3cm}}|
@{\hspace{0.3cm}}c@{\hspace{0.3cm}}|@{\hspace{0.3cm}}c@{\hspace{0.3cm}}
|@{\hspace{0.3cm}}c@{\hspace{0.3cm}}|@{\hspace{0.3cm}}c@{\hspace{0.3cm}}
|} \hline
$ $ & $z=120$ & $z=300$ & $z=500$ & $z=10^3$ & $z=3\times10^3$ & $z=10^4$ & $z=10^5$  \\
\hline
$\chi^2(B0.1A7)$ & $67.13$ & $65.78$ & $64.8$ & $64.49$ & $64.66$ & $64.87$ & $64.99$   \\
\hline
$\chi^2(B0.1A5)$ & $69.68$ & $65.75$ & $64.76$ & $64.49$ & $64.68$ & $64.87$ & $64.98$   \\
\hline
$\chi^2(B0.1A1)$ & $74.51$ & $65.49$ & $64.6$ & $64.50$ & $64.75$ & $64.91$ & $64.99$   \\
\hline
$\chi^2(B0.1A0.1)$ & $73.54$ & $65.21$ & $64.5$ & $64.51$ & $64.80$ & $64.95$ & $64.99$   \\
\hline
\end{tabular}
\caption{$\chi^2$ values for $B=0.1$ and $A=7, 5, 1, 0.1$, using
WMAP1 data.} \label{B0.1wmap1}
\end{table}

\begin{figure}[htp]
\centering
\includegraphics[width=0.45\textwidth]{fig2a.eps}
\includegraphics[width=0.45\textwidth]{fig2b.eps}
\end{figure}
\begin{figure}[htp]
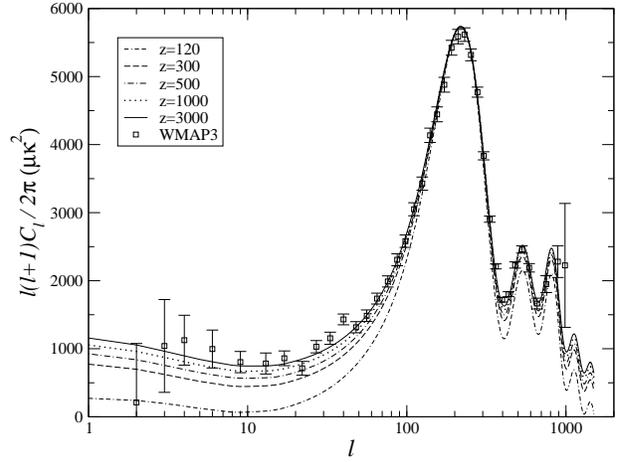

\centering
\includegraphics[width=0.45\textwidth]{fig2c.eps}
\includegraphics[width=0.45\textwidth]{fig2d.eps}
\caption{CMB anisotropy power spectra for $B=1$ and $A=7, 5, 1,
0.1$. In each panel, the curves are induced by the $P_k$ with
$z\equiv k_{0.05}/\Lambda=120, 300, 500, 1000, 3000$ corresponding
to $0.05\textrm{Mpc}^{-1}$. The $z=3000$ curve almost overlaps with
that of the $\Lambda$CDM model with a scale-invariant power
spectrum. We normalize all the anisotropy spectra at the first
Doppler peak. Also shown are the three-year WMAP data including
error bars.} \label{fig2}
\end{figure}

\begin{figure}[htp]
\centering
\includegraphics[width=0.45\textwidth]{fig3a.eps}
\includegraphics[width=0.45\textwidth]{fig3b.eps}
\end{figure}
\begin{figure}[htp]
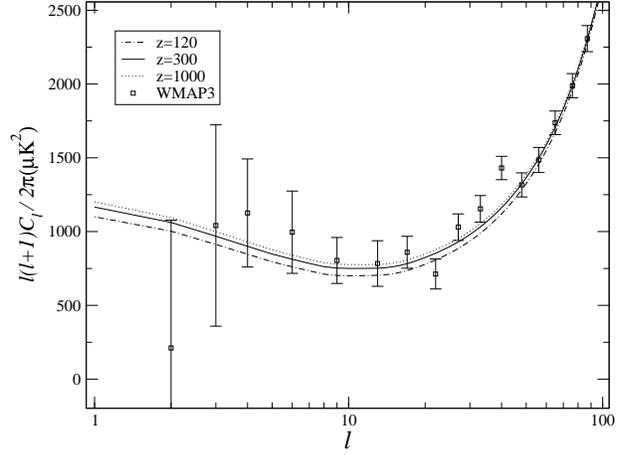

\includegraphics[width=0.45\textwidth]{fig3c.eps}
\includegraphics[width=0.45\textwidth]{fig3d.eps}
\caption{The same as in Fig.~\ref{fig2} but for $B= 0.1$ and $A= 7,
5, 1, 0.1$.} \label{fig3}
\end{figure}

\end{document}